\documentclass[12pt]{article}

\usepackage{array} 
\usepackage{epsfig}
\usepackage{amssymb}
\usepackage{amsmath}                          % roman hinzugefuegt
\usepackage{graphics,graphpap}

\renewcommand{\thefootnote}{\fnsymbol{footnote}}

\begin{document}

%%%%%%%%%% Title page
\begin{titlepage}
\begin{flushright}
\begin{tabular}{l}
IPPP/05/37\\
DCPT/05/74\\
TPI-MINN-05/22\\
\end{tabular}
\end{flushright}
\vskip1.5cm
\begin{center}
   {\Large \bf \boldmath $|V_{ub}|$ and Constraints on the
     Leading-Twist\\[5pt]
Pion   Distribution Amplitude from $B\to\pi \ell\nu$}
    \vskip1.3cm {\sc
Patricia Ball\footnote{Patricia.Ball@durham.ac.uk}$^{,1}$ and 
    Roman Zwicky\footnote{zwicky@physics.umn.edu}}$^{,2}$
  \vskip0.5cm
        $^1$ IPPP, Department of Physics, 
University of Durham, Durham DH1 3LE, UK \\
\vskip0.4cm 
$^2$ William I. Fine Theoretical Physics Institute, \\ University of
Minnesota, Minneapolis, MN 55455, USA  
\vskip2cm

%{\em Version of \today}

\vskip3cm

{\large\bf Abstract:\\[10pt]} \parbox[t]{\textwidth}{Using new
  experimental data on the leptonic mass spectrum of $B\to\pi \ell\nu$,
  we simultaneously determine $|V_{ub}|$ and constrain $a_2^\pi$ and 
$a_4^\pi$, the first two
   Gegenbauer moments of the pion's
  leading-twist distribution amplitude. We find $|V_{ub}|
  = (3.2\pm 0.1\pm 0.1\pm 0.3)\times 10^{-3}$, where the first error is
  experimental, the second comes from the shape of the form factor in $q^2$ 
  and the third is a 8\% uncertainty from the 
  normalisation of the form
  factor. We also find $a_2^\pi(1\,{\rm
  GeV})=0.19\pm{0.19}$ and $a_4^\pi(1\,{\rm
  GeV})\geq -0.07$. 
}

\vfill

{\em submitted to Physics Letters B}
\end{center}
\end{titlepage}

\setcounter{footnote}{0}
\renewcommand{\thefootnote}{\arabic{footnote}}

\newpage

\noindent
$|V_{ub}|$ is one of the least well-known elements of the CKM
quark-mixing matrix. A more precise determination of this parameter
will not only greatly improve the constraints on the unitarity
triangle, but also provide a stringent test of the CKM mechanism of
flavour structure and CP violation. In this letter, we determine
$|V_{ub}|$ from the exclusive semileptonic decay $B\to\pi\ell\nu$,
based on the invariant lepton-mass spectrum 
recently reported by BaBar \cite{data} and the
light-cone sum rule calculations of the relevant form factor
in Ref.~\cite{BZ04}. 

The hadronic matrix element relevant for $B\to\pi\ell\nu$ is given by
\begin{equation}
\langle \pi(p_\pi)| \bar u \gamma_\mu b | B(p_B)\rangle = \left(
p_B+p_\pi - q\,\frac{m_B^2-m_\pi^2}{q^2}\right)_\mu f_+(q^2) + 
 \frac{m_B^2-m_\pi^2}{q^2}\, q_\mu \,f_0(q^2),
\end{equation}
where the form factors $f_{+,0}$ depend on $q^2\equiv(p_B-p_\pi)^2$, 
the invariant mass
of the lepton-pair, with $0\,$GeV$^2\leq q^2\leq
(m_B-m_\pi)^2=26.4\,$GeV$^2$. $f_+$ is the dominant form factor,
i.e.\ the only one needed for calculating the spectrum in $q^2$,
\begin{equation}
\frac{d\Gamma}{dq^2}\,(B^0\to \pi^- \ell^+ \nu_\ell) = \frac{G_F^2
  |V_{ub}|^2}{ 192 \pi^3 m_B^3}\,\lambda^{3/2}(q^2) |f_+(q^2)|^2
\end{equation}
for massless leptons; $\lambda(q^2) = (m_B^2+m_\pi^2-q^2)^2 - 4 m_B^2
m_\pi^2$ is the usual phase-space factor. The determination of
$|V_{ub}|$ from $B\to\pi\ell\nu$ requires theoretical input on $f_+$,
which has been the subject of many a calculation 
using various methods, in particular quark models
\cite{QM}, QCD sum rules on the light-cone (LCSRs)
\cite{BZ04,BZ01,otherLCSR} and lattice simulations \cite{lattice}. 
The challenge for theory is
twofold: the region of applicability
of theoretical calculations is, in most cases, restricted to part of
the full physical phase-space; a calculation of $f_+$ for, say, small
values of $q^2$ is, however,
not sufficient, as experimental data on the decay
spectrum are still very scarce, so that any meaningful extraction of
$|V_{ub}|$  necessitates the extrapolation of the form factor to all
$q^2$. LCSR calculations, for instance, are valid for large pion
momentum, which translates into small to moderate
$q^2\,$\raisebox{-3pt}{$\stackrel{<}{\sim}$}$\,14\,$GeV$^2$,
whereas lattice calculations are restricted to small pion momentum, 
corresponding to large
$q^2\,$\raisebox{-3pt}{$\stackrel{>}{\sim}$}$\,15\,$GeV$^2$. 
Extrapolations rely either on a model for the
$q^2$-dependence of $f_+$, like vector meson dominance or the
parametrisation advocated by Becirevic and Kaidalov \cite{BK},
or dispersive bounds
on the form factor, which have been studied for instance 
in Ref.~\cite{dispersion}.
The BaBar collaboration has measured the spectral decay distribution in
5 bins in $q^2$ \cite{data}, which is a significant improvement over
previous results reported for 3 bins \cite{prevdata}, and allows one, for the
first time, to assess the validity of various 
parametrisations of the $q^2$-dependence of $f_+$ like
\begin{itemize}
\item vector meson dominance (VMD);
\item the parametrisation of Becirevic and Kaidalov (BK) \cite{BK};
\item the extended BK parametrisation used by Ball and Zwicky
  (BZ) \cite{BZ04}.
\end{itemize}
All these parametrisations 
can be motivated from the exact representation of $f_+$
in terms of a dispersion relation,
\begin{equation}\label{eq:disper}
f_+(q^2) = \frac{{\rm Res}_{q^2=m_{B^*}^2} f_+(q^2)}{q^2-m_{B^*}^2} +
\frac{1}{\pi} \,\int_{(m_B+m_\pi)^2}^\infty dt\,\frac{{\rm
    Im}\,f_+(t)}{t-q^2-i \epsilon}\,,
\end{equation}
where $m_{B^*}=5.325\,$GeV is the mass of the $B^*(1^-)$ meson which
induces a pole below the lowest multiparticle threshold at $q^2=(m_B+m_\pi)^2$.
Vector meson dominance assumes that the form factor is dominated, for
all physical $q^2$, by the first term in (\ref{eq:disper}):
\begin{equation}\label{para:VMD}
\left.f_+(q^2)\right|_{\rm VMD} =
\frac{f_+(0)}{1-q^2/m_{B^*}^2}\,;
\end{equation}
$f_+(0)$ is the only free parameter of the VMD parametrisation.
Becirevic and Kaidalov suggested as an alternative parametrisation 
the replacement of the second term in
(\ref{eq:disper}) by an effective
pole at higher mass,
\begin{equation}\label{para:BZ}
\left.f_+(q^2)\right|_{{\rm BZ}} = 
\frac{r_1}{1-q^2/m_{B^*}^2} + \frac{r_2}{1-\alpha\, q^2/m_{B^*}^2}
\equiv \frac{f_+(0)}{1-q^2/m_{B^*}^2} + 
\frac{r\, q^2/m_{B^*}^2}{(1-q^2/m_{B^*}^2)(1-\alpha\,q^2/m_{B^*}^2)}
\end{equation}
with the three parameters $(r_1,r_2,\alpha)$ or $(f_+(0),r,\alpha)$,
which are related by
\begin{equation}\label{redef}
f_+(0) = r_1+r_2,\qquad r = r_2 (\alpha-1),
\end{equation}
where $0<\alpha<1$ parameterises the position of the effective pole.
This parametrisation was used by BZ to describe the results
from LCSRs \cite{BZ04}, whereas Becirevic and Kaidalov,
faced with the challenge to fit three independent parameters to
lattice data with limited accuracy, implemented the additional
constraint $r\equiv \alpha f_+(0)$ 
motivated from heavy quark expansion, and obtained the
following expression in terms of two parameters, $(c_B,\alpha)$ or 
$(f_+(0),\alpha)$:
\begin{equation}\label{para:BK}
\left.f_+(q^2)\right|_{{\rm BK}} =
\frac{c_B(1-\alpha)}{(1-q^2/m_{B^*}^2)(1-\alpha\,q^2/m_{B^*}^2)}
\equiv \frac{f_+(0)}{ (1-q^2/m_{B^*}^2)(1-\alpha\,q^2/m_{B^*}^2)}\,.
\end{equation}

The BaBar collaboration has measured the integrated $q^2$-spectrum of
$B\to\pi \ell\nu$ in five bins in $q^2$ \cite{data}, which allows one
to confront the above parametrisations with experiment. Fitting the data
by the VMD formula (\ref{para:VMD}), we 
find $\chi^2 = 9.2/4$ d.o.f. 
The BK parametrisation (\ref{para:BK}) 
fits the data with $\chi^2 = 3.5/3$ d.o.f.\ and $\alpha = 0.61\pm
0.09$. Fitting the BZ parametrisation (\ref{para:BZ}) is slightly more
subtle, as the data prefer $\alpha\to 1$ or even larger, which is
outside the allowed parameter-space.\footnote{For $\alpha=1$ the
rescaling (\ref{redef}) is no longer valid.}
Bounding $\alpha\leq 1$, we find
a minimum $\chi^2 = 1.65/2\,$d.o.f.\  and $r/f_+(0) = 0.24\pm 0.08$, 
$\alpha=1.00^{+0}_{-0.15}$. 
While this implies that the VMD parametrisation is 
disfavoured,\footnote{Actually VMD is disfavoured 
not only from the experimental point of
  view, but also from the theoretical one, as the values of $f_+(0)$
  do not agree with indepedent determinations of the residue of the
  $B^*$ pole, see the discussion in
  Sec.~4 of Ref.~\cite{BZ04}.} 
both BK and BZ are viable parametrisations. 
Motivated by these results, we formulate the following strategy for
extracting $|V_{ub}|$ from the data: we
\begin{itemize}
\item calculate $f_+$ from LCSRs for values of $q^2$ where the
method is applicable;
\item extrapolate the results to all $q^2$ using 
the experimentally favoured BK and  BZ parametrisations;
\item use the experimental
information on the spectrum to constrain the input parameters
of the LCSRs, in particular the leading-twist $\pi$ distribution
amplitude;
\item determine $|V_{ub}|$ from the total branching ratio
using the BK and BZ parametrisations of $f_+$.
\end{itemize}

\begin{figure}
$$\epsfxsize=0.5\textwidth\epsffile{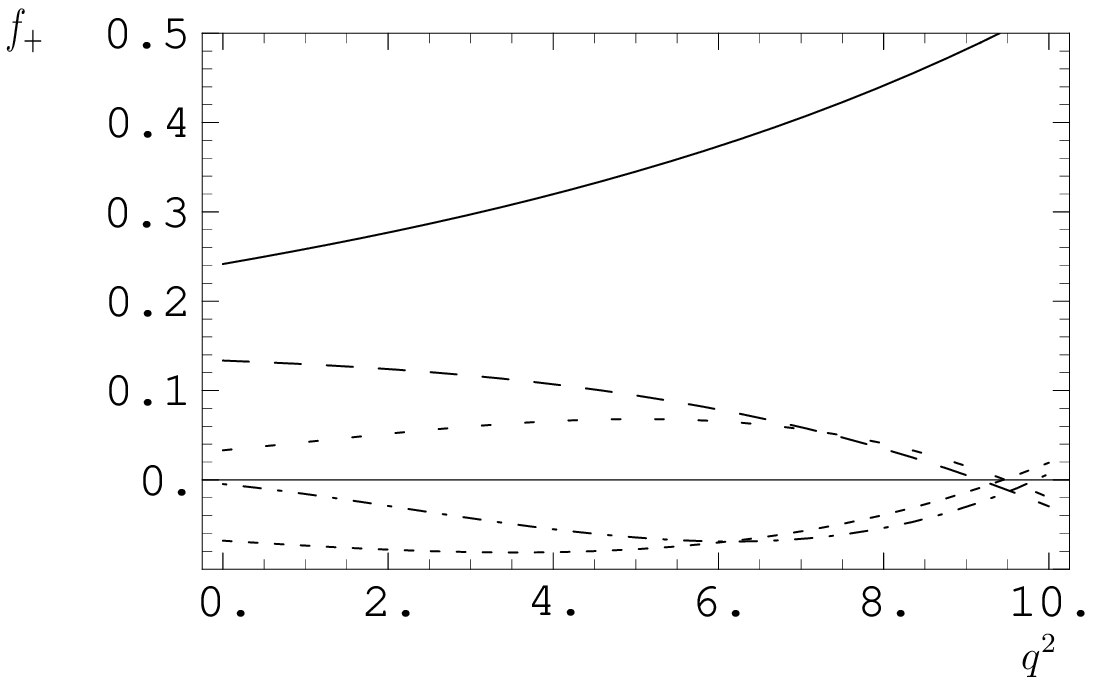}$$
\caption[]{Respective size of different contributions in $a_n^\pi$ to
  $f_+(q^2)$ as a function of $q^2$ for central values of the LCSR
  input parameters.
Solid line: $a_n^\pi=0$. Remaining lines from top to
  bottom (at $q^2=0$): contributions proportional to $a_2^\pi$,
  $a_6^\pi$, $a_8^\pi$, $a_{4}^\pi$.}\label{fig:shape}
\end{figure}

Let us start with the calculation of $f_+$. 
In Ref.~\cite{BZ04}, we have presented a comprehensive analysis of
$B\to(\pi,K,\eta)$ decay form factors calculated from QCD sum rules on
  the light-cone, to $O(\alpha_s)$ accuracy for twist-2 and the
  dominant twist-3 contributions; earlier analyses can be found in
  Refs.~\cite{BZ01,otherLCSR}. We refer to these papers for an
   explanation of the method. The main theoretical 
uncertainty of these analyses comes from the pion's leading-twist light-cone 
distribution amplitude (DA) $\phi_\pi$; 
other sources of uncertainty include the $b$
quark mass, the quark condensate and sum rule specific
parameters (Borel parameter and continuum threshold). The resulting 
total uncertainty of $f_+$ is between 10\% and 13\%
\cite{BZ04}. Whereas the other parameters mainly determine the
normalisation of the form factor, $\phi_\pi$
affects also and in particular the $q^2$-dependence and hence
can be constrained from the measured spectrum. 
The DA is usually expressed in terms of its conformal expansion,
\begin{equation}\label{eq:conformal}
\phi_\pi(u,\mu) = 6u(1-u) \left( 1 + \sum_{n=1}^\infty a_{2n}^\pi(\mu)
C^{3/2}_{2n}(2u-1)\right),
\end{equation}
where $u$ is the momentum fraction of the quark in the $\pi$ and runs
from 0 to 1. The $C^{3/2}_n$ are Gegenbauer polynomials and $a_n$, the
so-called Gegenbauer moments, are hadronic parameters which depend on
the factorisation scale $\mu$. The respective contributions of
$a_{n\leq 8}^\pi$ to $f_+$ are shown in 
Fig.~\ref{fig:shape}, for a typical choice of input parameters. 
The plot reveals that the $q^2$-dependence of the form factor is mostly
sensititive to $a_2^\pi$ and only to a lesser extent to higher
Gegenbauer-moments, which agrees with the findings of
Ref.~\cite{talbot}. We hence decide against using the models for
$\phi_\pi$ proposed in Ref.~\cite{talbot}, but stick with the
expansion (\ref{eq:conformal}), which we truncate
after the contribution in $a_4^\pi$.
The values of the lowest lying Gegenbauer moments $a_{2,4}^\pi$ have
been constrained from various sources, cf.\ Ref.~\cite{BF,piDA,stefanis}, but 
still come with rather large uncertainties. A very conservative range
of allowed values consistent with all known constraints is
\begin{equation}\label{eq:a2a4}
0\leq a_2^\pi(1\,{\rm GeV})\leq 0.3, \qquad 
-0.15\leq a_4^\pi(1\,{\rm GeV})\leq 0.15.
\end{equation}
In this letter, we aim to constrain $a^\pi_{2,4}$ from experimental data
within the above range.

We obtain values for
$f_+(q^2)$ in dependence on $a_2^\pi$, $a_4^\pi$ and $m_b$ 
using the following criteria for
the evaluation of the LCSRs:
\begin{itemize}
\item we calculate $f_+(q^2)$ as a function of the Borel parameter
  $M^2$ and the continuum threshold $s_0$ for two different values of $m_b$,
  $m_b\in\{4.7,4.8\}\,$GeV, five different values of $q^2$, $q^2\in\{0,
  2.5,5,7.5,10\}\,$GeV$^2$, and 16 different values of
  $(a_2^\pi(1\,{\rm GeV}),a_4^\pi(1\,{\rm GeV}))$: $a_2^\pi\in\{
  0,0.1,0.2,0.3\}$, $a_4^\pi\in\{-0.15,-0.05,0.05,0.15\}$;\footnote{In
  the actual calculation, $a_{2,4}^\pi$ are scaled up to the factorisation
  scale $\mu =\sqrt{m_B^2- m_b^2} \approx 2.2\,$GeV using NLO
  evaluation.} we interpolate the results in $a_2^\pi$ and
  $a_4^\pi$ in order to obtain $f_+$ as a smooth function of these
  parameters;  
\item for each value of the input parameters, 
  we determine $f_+$ at the minimum in the Borel parameter
  $M^2$, which implies that $M^2$ becomes mildly dependent on $q^2$. 
  As discussed in Ref.~\cite{BZ04},
  this procedure ensures that a LCSR for $m_B$, obtained from the derivative of
  the LCSR for $f_+$  in $M^2$, yields the physical value $m_B = 5.28\,$GeV;
\item we choose the continuum threshold $s_0$ in such a way that the
  continuum contribution to the LCSR is constant for all
  $q^2$; this implies that also $s_0$ becomes (mildly) dependent
  on $q^2$. For each value of the input parameters, we calculate $f_+$
  for three different values of the continuum contribution, 15\%, 20\%
  and 25\%;
\item the LCSR actually yields $f_Bf_+$, $f_B$ being the leptonic
  decay constant of the $B$ meson; in order to extract $f_+$, we
  divide the LCSR by $f_B$ as calculated from a QCD sum rule to the same
  accuracy in $\alpha_s$. 
\end{itemize}
Some of these criteria differ from those applied in
Ref.~\cite{BZ04}. We chose to modify the criteria used in our
previous work since we focus, in this letter,
on the dependence of the form factor on $a^\pi_{2,4}$, which can 
be meaningfully determined only if $f_+$ is calculated using
exactly the same criteria for all values of 
$q^2$, $m_b$, $a_2^\pi$ and $a_4^\pi$. It is for this reason
that we require the continuum contribution to be the same for all
input parameters. The drawback of this
procedure is that the sum rule specific parameters $M^2$ and $s_0$
both become dependent on $q^2$, so that, in order to keep the calculational
effort at a manageable level, we have to restrict ourselves to a few
points in $q^2$.
For each value of $f_+$ we 
calculate the theoretical uncertainty by varying
\begin{itemize}
\item $M^2$ by 40\% around the central value;
\item $s_0$ by $\pm 1\,$GeV$^2$;
\item the central value 20\% of the continuum contribution between
  15\% and 25\%;
\item the central value of the quark condensate, $-\langle \bar q q
  \rangle (1\,{\rm GeV}) = (0.24\,{\rm GeV})^3$, between $((0.24\pm
  0.01)\,{\rm GeV})^3$.
\end{itemize}
The above ranges of sum rule parameters are rather conservative and
account for the ``systematic'' uncertainty of QCD sum rule calculations. 
All errors are added linearly, 
which yields a typical theoretical uncertainty of 8\%. 
\begin{figure}
$$\epsfxsize=0.5\textwidth\epsffile{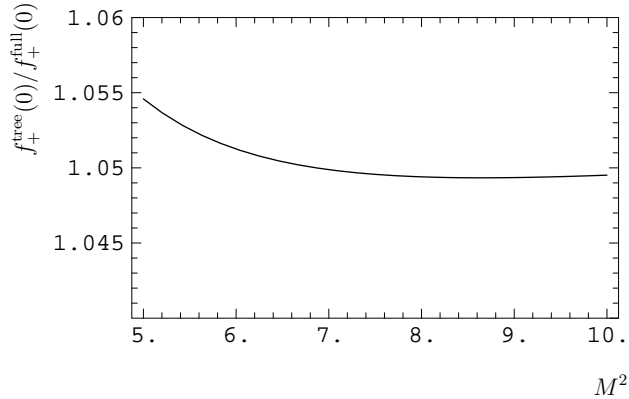}$$
\caption[]{$f_+^{\rm tree}(0)/f_+^{\rm full}(0)$
  as a function of the Borel parameter $M^2$ for $a_2^\pi = 0.1$,
  $a_4^\pi = 0$, $m_b = 4.8\,$GeV, $s_0^{\rm tree} = 35.25\,$GeV$^2$,
  $s_0^{\rm full} = 34.42\,$GeV$^2$, which corresponds to 20\%
  continuum contribution. The ratio of the decay constants
  is $f_B^{\rm full}/f_B^{\rm tree} = 1.26$.}\label{fig:norm}
\end{figure}
At this point we would like to comment on the treatment of $f_B$.
The motivation for calculating $f_B$ from a QCD sum rule instead of
using, for instance, the current world average from lattice
calculations \cite{fB_latt}, is that (a) 
$f_B$ receives large $O(\alpha_s)$ and $O(\alpha_s^2)$ corrections
from gluon-exchange diagrams that also enter the LCSR for
$f_Bf_+$ and (b) $f_B$ is very sensitive to $m_b$. By dividing the
LCSR for $f_Bf_+$ by $f_B$ obtained 
from a QCD sum rule to the same accuracy in $\alpha_s$, and using the
same value of $m_b$, one  expects
those large contributions to cancel and to reduce the sensitivity to
the value of $m_b$. One can check the
extent to which the cancellation takes place by comparing
the results of our $O(\alpha_s)$ calculation with that at tree-level. 
\begin{table}
$$
\addtolength{\arraycolsep}{3pt}\renewcommand{\arraystretch}{1.2}
\begin{array}{c|cc}
\hline
m_b\,[{\rm GeV}] & 4.7 & 4.8\\\hline
f_B\,[{\rm MeV}] & 191\pm 6 & 164 \pm 4\\\hline
\end{array}
\addtolength{\arraycolsep}{-3pt}\renewcommand{\arraystretch}{1}
$$
\caption[]{$f_B$ from a QCD sum rule to $O(\alpha_s)$ accuracy; the
  error is obtained by varying $(M^2)_{f_B}$ by 40\%  and
  $(s_0)_{f_B}$ by $\pm 1\,$GeV$^2$. The current world average
  from lattice calculations is $f_B = (189\pm 27)\,$MeV 
\cite{fB_latt}.}\label{tab:fB}
\end{table}
We calculate $f_+^{\rm tree}(0)\equiv(f_Bf_+(0))^{\rm tree}/(f_B)^{\rm
  tree}$ for $m_b = 4.8\,$GeV, $a_2^\pi = 0.1$, $a_4^\pi = 0$
using the same criteria as for the full form factor $f_+^{\rm full}(0)$ 
including
$O(\alpha_s)$ corrections. In Fig.~\ref{fig:norm} we plot the ratio
$f_+^{\rm tree}(0)/f_+^{\rm full}(0)$ for the respective optimum
values $s_0$ as a
function of $M^2$. The minima in $M^2$ are around $7\,$GeV$^2$. The
ratio is nearly constant $\sim 1.05$, whereas the ratio of the decay
constants, $f_B^{\rm full}/f_B^{\rm tree}$ is 1.26. 
This means that there is indeed a strong cancellation between
the radiative corrections to the LCSR for $f_Bf_+(0)$ and the QCD sum
rule for $f_B$. We have checked that similar cancellations also occur
for other values of $m_b$ and nonzero $q^2$. 
Nonetheless there is a residual uncertainty due to the treatment of
$f_B$ which we estimate to be about half the difference between the value of
$f_B$ calculated from a QCD sum rule to $O(\alpha_s)$ accuracy
and the central lattice value, i.e.\ about
5\%. The QCD sum rule results for $f_B$ are given, with errors, 
in Tab.~\ref{tab:fB}.
Adding these two errors linealy, we obtain an uncertainty of
  $f_B$ of 8\% which is independent of $q^2$ and translates into a 8\%
  uncertainty of the normalisation of $f_+$, which we treat
  separately from the error of $f_Bf_+$ calculated as described
  above.\footnote{As the quark condensate is a common input parameter
  for both $f_Bf_+$ and $f_B$, the effect of its variation is included
  only once, in the uncertainty of $f_Bf_+$.}

\begin{figure}
$$\epsfysize=0.18\textheight\epsffile{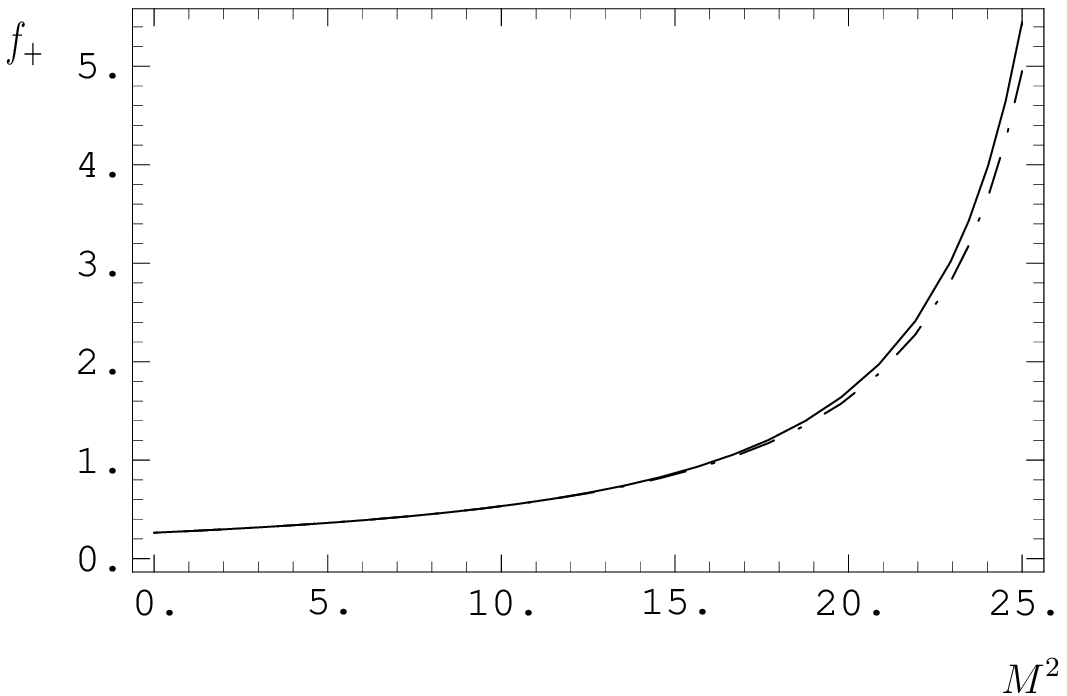}\quad
\epsfysize=0.18\textheight\epsffile{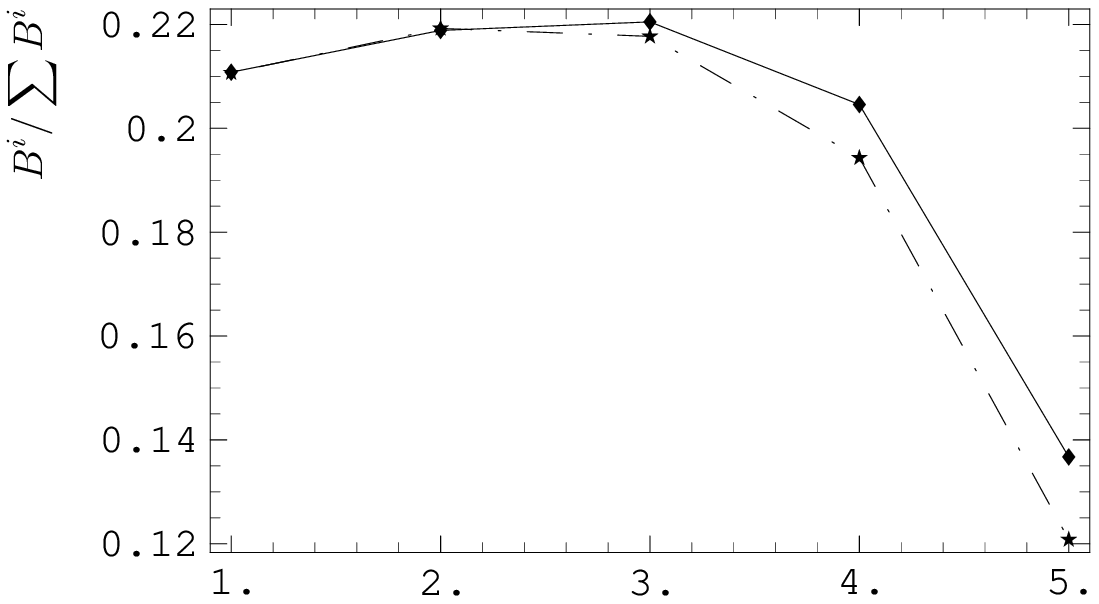}
$$
\caption[]{Left: $f_+(q^2)$ as a function of $q^2$ 
for $m_b=4.8\,$GeV, $a_2^\pi = 0.115$ and
  $a_4^\pi=-0.015$ fitted to the BK parametrisation (\ref{para:BK})
  (solid line) and to the BZ parametrisation (\ref{para:BZ})
  (dot-dashed line). Right: corresponding integrated spectra in the
five $q^2$-bins used by BaBar.}\label{fig:BKBZ}
\end{figure}

The next task is to compare the form factor predictions to data and
to determine best-fit values of $a_{2,4}^\pi$, using the
experimentally favoured BK and BZ parametrisations in order to extrapolate
the LCSR results to all physical $q^2$. It turns out that the 
LCSR results are actually described
extremely well by the BK and BZ parametrisations, to within
better than 0.5\%, as already noted in Ref.~\cite{BZ04}. 
In Fig.~\ref{fig:BKBZ} we show the difference between
the BK and the BZ fit, both for the form factor and the integrated
$q^2$-spectra in the $i$th bin, $5(i-1){\rm GeV}^2 \leq q^2 \leq 5i{\rm
  GeV}^2$, $1\leq i\leq 5$. The form factors start to noticeably
deviate only for very large $q^2$. For the fit of the experimental spectrum, 
we treat the experimental
errors as uncorrelated, but allow for a correlation of theory errors
and perform the least-$\chi^2$ fits using the following $\chi^2$ function and 
error matrix $E$:
\begin{eqnarray*}
\chi^2 &=& \sum_{i,j=1}^5 (B_i^{\rm
  th} - B_i^{\rm exp}) (E^{-1})_{ij} (B_j^{\rm
  th} - B_j^{\rm exp}),\\
E_{ij} &=& (\sigma_i^{\rm exp})^2 \delta_{ij} + (\Delta B_i^{\rm
  th})^2 \delta_{ij} + {\cal C}^2 (\Delta B_i^{\rm th}) (\Delta
  B_j^{\rm th}) (1-\delta_{ij}),
\end{eqnarray*}
where $B_i$ is the partial branching fractions $i$th $q^2$-bin, 
$\Delta B_i^{\rm th}$ the corresponding theory error 
(without the error of the overall
  normalisation of $f_+$) and ${\cal C}$ is the correlation of
  theory errors
  which we vary within $0\leq {\cal C}\leq 1$.

\begin{figure}[tb]
$$\epsfxsize=0.45\textwidth\epsffile{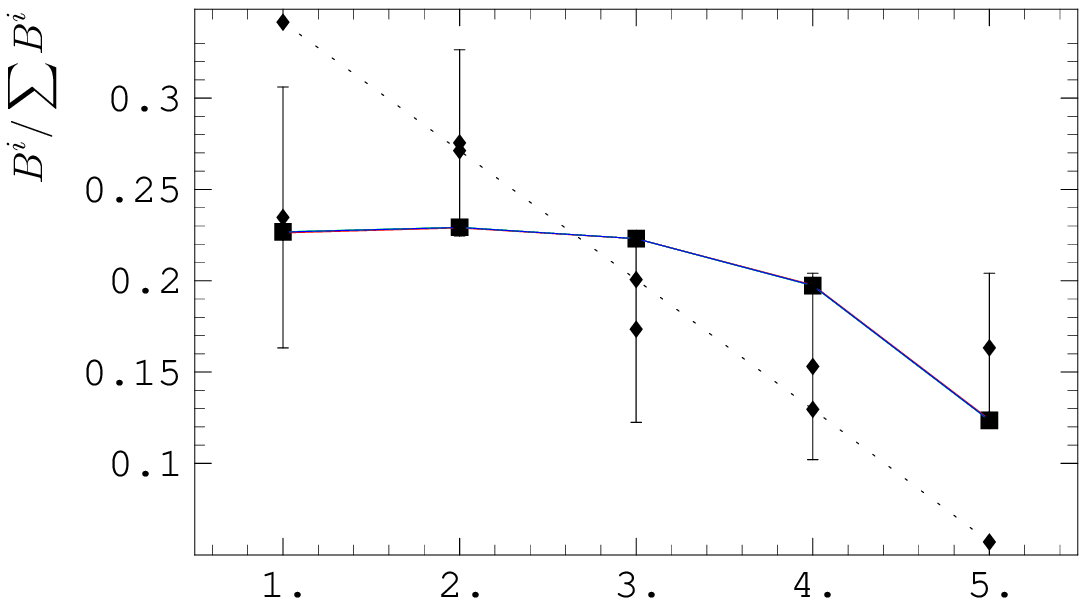}\qquad\epsfxsize=0.45\textwidth\epsffile{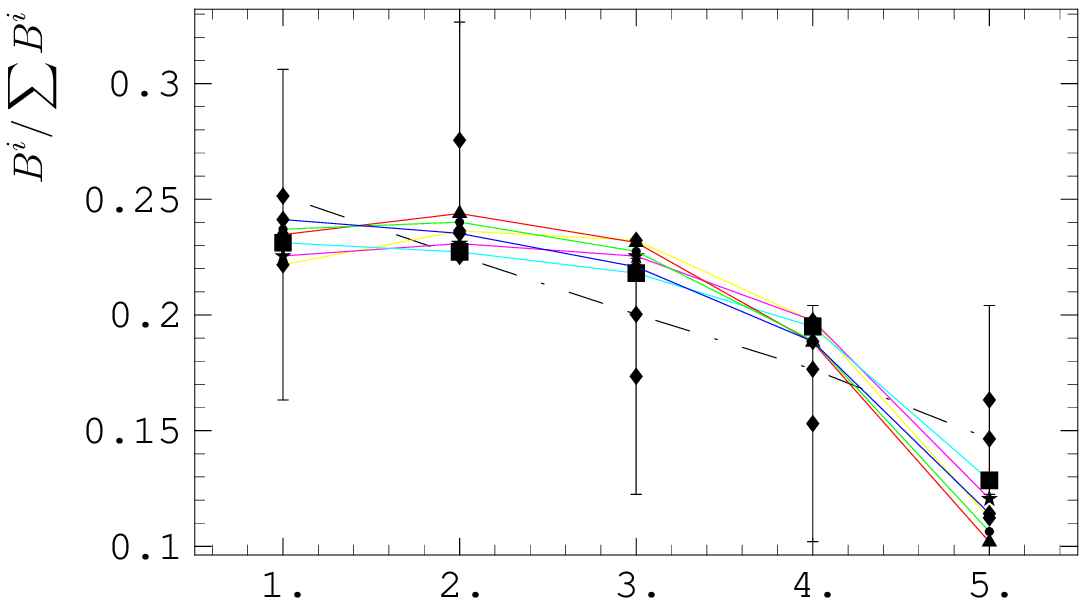}$$
\caption[]{(Colour online) Best fit results for the partially
  integrated differential decay rates (solid lines) and
  experimental data (partially integrated spectra in $q^2$-bins 1 to
  5). 
Left: BK parametrisation of the form factor,
  right: BZ parametrisation. Solid lines: set of input parameters as in Tabs.~\ref{tab:1}
  and \ref{tab:2}. Dotted line: VMD parametrisation,
  Eq.~(\ref{para:VMD}). Dash-dotted line: best BZ fit to experimental data.}\label{fig:3}
\end{figure}

We first study the BK
parametrisation which features one parameter that can be
determined from the experimental spectrum: $\alpha=0.61\pm 0.09$. As $f_+$ depends
on actually three parameters, $a_2^\pi$, $a_4^\pi$ and $m_b$, only one
of them can be constrained. In Tab.~\ref{tab:1} we give the best-fit
values of $a_2^\pi$ for $a_4^\pi$ and $m_b$ fixed. The table reveals
that it is indeed possible to reproduce the experimental central value
of $\alpha$ for any given $a_4^\pi$ and $m_b$. It is also obvious that
the impact of the precise value of $a_4^\pi$ is much smaller than that
of $m_b$. Averaging over $a_4^\pi$ within $[-0.15,0.15]$ and over
$m_b$, we find
\begin{equation}\label{a2BK}
\left.a_2^\pi(1\,{\rm GeV})\right|_{\rm BK} = 0.14^{+0.18}_{-0.20}\,.
\end{equation}
We would like
to stress that this is a completely new determination of
$a_2^\pi$ and agrees very well with other determinations of this
parameter \cite{BF,piDA,stefanis}.
What is truly remarkable, however, is that, despite 
different best-fit values of $a_2^\pi$, the resulting values of
$f_+(0)$, and hence $|V_{ub}|$, agree within 3\%. That is: the theory
error due to $m_b$ and $a_{2,4}^\pi$ gets largely diminished by the 
constraints on the spectrum.
Using the parameter sets in Tab.~\ref{tab:1}, we plot, in
Fig.~\ref{fig:3}, the partially
integrated spectra, normalised to the full branching ratio, together
with the experimental data. All six parameter sets produce nearly the same
curve which coincides with the best experimental fit using the BK 
parametrisation. Using the average value of the branching ratio as
given by HFAG \cite{HFAG}, $B(B^0\to\pi^-\ell^+\nu_\ell) = (1.36\pm
0.11)\times 10^{-4}$, we obtain
\begin{equation}\label{VubBK}
|V_{ub}|_{\rm BK} = (3.2\pm 0.1\pm 0.1 \pm 0.3)\times 10^{-3},
\end{equation}
where the first error is experimental, the second comes 
from the uncertainty in the shape, due to the spread of 
values of $a_{2,4}^\pi$, and the third is from the normalisation of
the form factor.

\begin{table}[tb]
$$
\addtolength{\arraycolsep}{4pt}\renewcommand{\arraystretch}{1.2}
\begin{array}{cr|rrrr}
\hline
m_b & \multicolumn{1}{c|}{a_4^\pi} & \multicolumn{1}{c}{a_2^\pi} & 
\multicolumn{1}{c}{f_+(0)} & \multicolumn{1}{c}{\alpha} & 
\multicolumn{1}{c}{|V_{ub}|}\\\hline
4.8\,{\rm GeV} & -0.15 & 0.19\pm 0.13 & 0.277 & 0.61  & 3.17\times 10^{-3}\\
& 0 & 0.18\pm 0.14 & 0.272 & 0.61  & 3.23\times 10^{-3}\\
& 0.15 & 0.16\pm 0.15 & 0.268 & 0.61 & 3.29\times 10^{-3}\\
4.7\,{\rm GeV} & -0.15 & 0.08\pm 0.13 & 0.274 & 0.61 & 3.21\times 10^{-3}\\
& 0 & 0.11\pm 0.14 & 0.272 & 0.61 & 3.23\times 10^{-3}\\
& 0.15 & 0.14\pm 0.16 & 0.271 & 0.61 & 3.25\times 10^{-3}\\\hline
\end{array}
\addtolength{\arraycolsep}{-4pt}
$$
\caption[]{$a_2^\pi(1\,{\rm GeV})$ from the fit to the shape of the
  experimental $q^2$-spectrum, for fixed $m_b$ and $a_4^\pi$ using the
  BK parametrisation (\ref{para:BK}). The
  table also gives the corresponding BK parameters $f_+(0)$ and
  $\alpha$ and the resulting central value of
  $|V_{ub}|$. $\chi^2_{\rm min} = 3.5/3\,$d.o.f.}\label{tab:1}
$$
\addtolength{\arraycolsep}{4pt}\renewcommand{\arraystretch}{1.2}
\begin{array}{cr|rrrrrr}
\hline
m_b & \multicolumn{1}{c|}{a_4^\pi} & \multicolumn{1}{c}{a_2^\pi} & 
\multicolumn{1}{c}{f_+(0)} & \multicolumn{1}{c}{r} & \multicolumn{1}{c}{\alpha} & 
\multicolumn{1}{c}{\chi^2} & \multicolumn{1}{c}{|V_{ub}|}\\\hline
4.8\,{\rm GeV} & 0  & 0.15 & 0.268 & 0.18 & 0.54 & 3.67 & 3.26\times 10^{-3}\\
& 0.15 & 0.23 & 0.278 &  0.14 & 0.73 & 3.06 & 3.22\times 10^{-3}\\
4.7\,{\rm GeV} & 0 & 0.10 & 0.270 & 0.18 & 0.52 & 3.71 & 3.26\times 10^{-3}\\
& 0.15 & 0.24 & 0.282 & 0.14 & 0.70 & 3.11 & 3.23\times 10^{-3}\\\hline
\end{array}
\addtolength{\arraycolsep}{-4pt}
$$
\caption[]{Ditto for the BZ parametrisation (\ref{para:BZ}). In
  addition we give the minimum $\chi^2$, for 3~d.o.f.,
for the fit of the form factor to the data.}\label{tab:2}
\end{table}

\begin{figure}[tb]
$$\epsfxsize=0.45\textwidth\epsffile{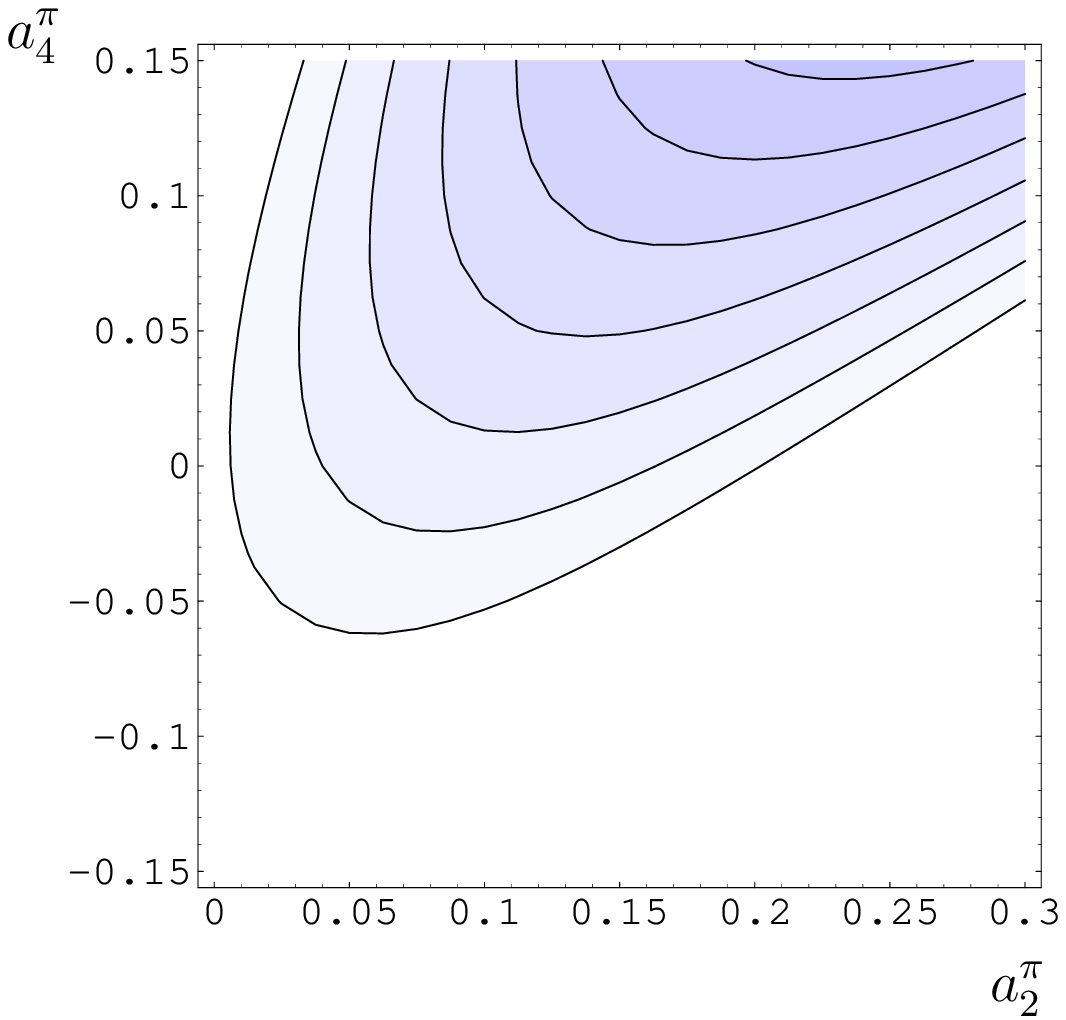}\quad
\epsfxsize=0.45\textwidth\epsffile{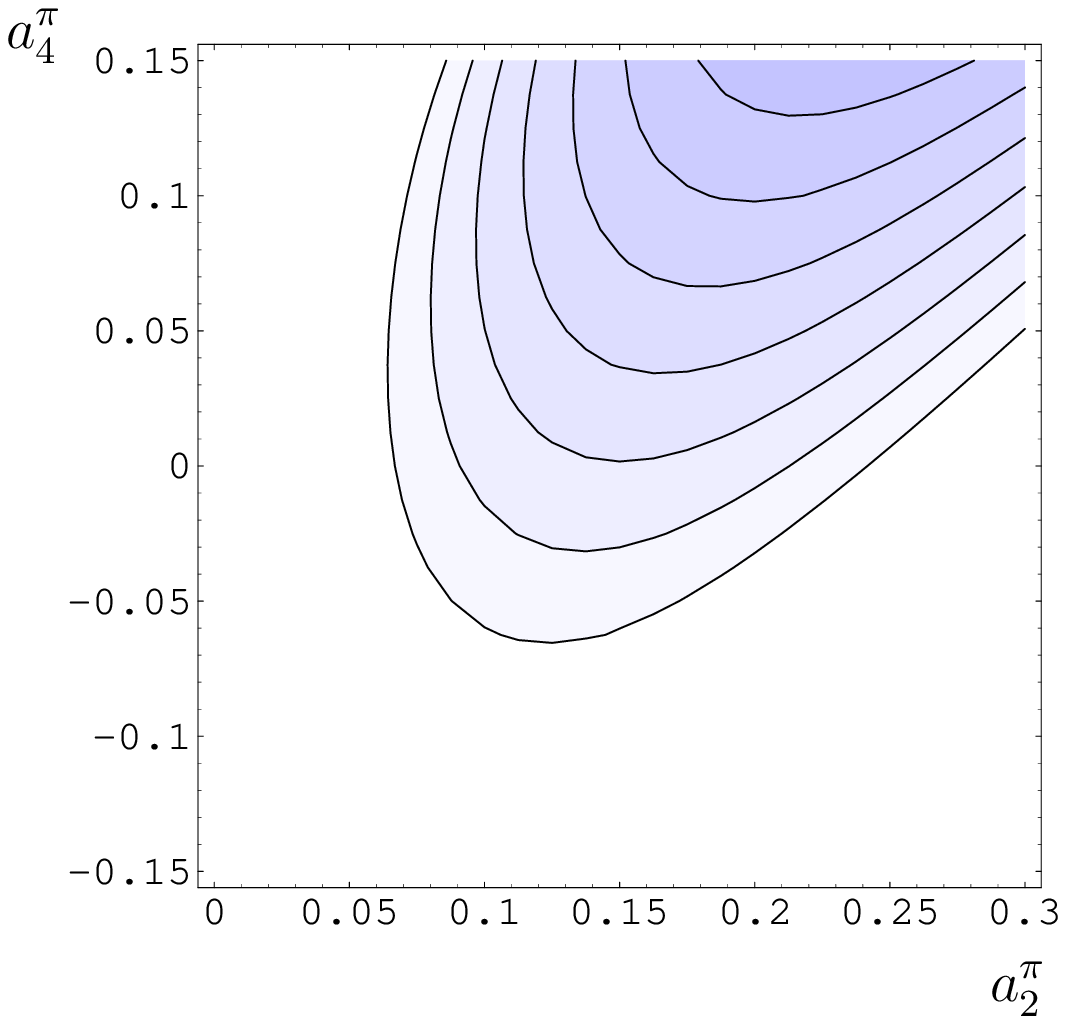}$$
\caption[]{(Colour online) 
Allowed values of $a_{2,4}^\pi$ for $m_b = 4.7\,{\rm GeV}$
  (left) and $m_b = 4.8\,{\rm GeV}$ (right). The minimum $\chi^2$ is
  3.06 and the contours include $(a_2^\pi,a_4^\pi)$ with $\chi^2\leq
  4.06$.}\label{fig:add}
\end{figure}

Let us now turn to the BZ parametrisation, which featurs two parameters
that can be determined from the shape of the spectrum, 
$r/f_+(0)=0.24\pm0.08$ and $\alpha=1.00^{+0}_{-0.15}$,
which allows one to constrain for instance
$a_{2}^\pi$ and $a_4^\pi$ in dependence on $m_b$. The resulting
constraints are shown in Fig.~\ref{fig:add}. The minimum $\chi^2$
for $(a_2^\pi,a_4^\pi)$ within the range specified in (\ref{eq:a2a4})
is reached for $a_2^\pi = 0.23$, $a_4^\pi = 0.15$, i.e.\ at the border of
the parameter space; fixing $a_4^\pi = 0.15$, the best-fit value of
$a_2^\pi$ is
\begin{equation}\label{a2BZ}
\left.a_2^\pi(1\,{\rm GeV})\right|_{\rm BZ}
 = 0.23\pm 0.15 \quad\mbox{for $a_4^\pi(1\,{\rm
 GeV})
 = 0.15$.}
\end{equation}
The contours shown in the figure include all
$(a_2^\pi,a_4^\pi)$ for which the fit of the corresponding form
factor to the data yields $\chi^2\leq \chi^2_{\rm min}+1=4.06$. We
immediately read off the following constraints:
\begin{equation}\label{a2a4const}
a_2^\pi(1\,{\rm GeV}) \geq 0,\qquad a_4^\pi(1\,{\rm GeV}) \geq -0.07.
\end{equation}
In Tab.~\ref{tab:2}, we give the fit results for $a_4^\pi=0$ and $0.15$.
The best-fit parameters do not agree, for $a_{2,4}^\pi$
within the range specified in (\ref{eq:a2a4}), with those favoured
by experiment. We have already pointed out earlier that the
experimentally favoured value $\alpha=1$ is actually outside the
theoretically allowed parameter space and is not supported by the
results of the LCSR calculation.
As before, we find that the different values for $a_{2,4}^\pi$ result in
nearly the same values of $|V_{ub}|$. In Fig.~\ref{fig:3} we plot the
best-fit values for the partially integrated branching fractions, obtained
from the parameter sets in Tab.~\ref{tab:2}, in comparison with  the
experimental results.
For $|V_{ub}|$, we find
\begin{equation}\label{VubBZ}
|V_{ub}|_{\rm BZ} = (3.2\pm 0.1\pm 0.1 \pm 0.3)\times 10^{-3}
\end{equation}
which agrees with (\ref{VubBK}).

Combining the results from the BK and the BZ analyses, we get the
following final result for $|V_{ub}|$: 
\begin{equation}\label{Vubtotal}
|V_{ub}| = (3.2\pm 0.1\pm 0.1 \pm 0.3)\times 10^{-3};
\end{equation}
the first error in $|V_{ub}|$ is experimental, the second comes from the shape
of the form factor and the third from the overall normalisation of $f_+$.
As for the constraints on $a_{2,4}^\pi$, our final results are
\begin{equation}\label{a2a4final}
a_2^\pi(1\,{\rm GeV}) = 0.19\pm 0.19, \quad a_4^\pi(1\,{\rm GeV}) \geq -0.07.
\end{equation}
The value of $a_2^\pi$ is the weighted average of (\ref{a2BK}) and
(\ref{a2BZ}). We would like to stress again that these values result from a
new and independent determination of these parameters, which is both
consistent with and 
complementary to the results found in Refs.~\cite{BF,piDA,stefanis}.

To summarize, we have discussed the constraints posed by recent
experimental data on the shape of the $B\to\pi$ decay form factor
$f_+$. We have found that both the BK parametrisation Eq.~(\ref{para:BK})
and the BZ parametrisation Eq.~(\ref{para:BZ}) can describe the data, 
whereas vector meson
dominance is disfavoured. We have calculated $f_+$ from QCD sum rules
on the light-cone for small to moderate $q^2$ and extrapolated the
form factor to
all physical $q^2$ using the BK and BZ parametrisations,
respectively. We have then used the experimental data to constrain the
pion distribution amplitude $\phi_\pi$ that enters the
calculation of $f_+$. We found that, although the best-fit values of
the Gegenbauer moments $a_{2,4}^\pi$  depend on $m_b$, the
resulting predictions for the form factors and the partial branching
fractions are largely independent of the input parameters. This is one
of the main results of this letter: the experimental information on
the shape of the spectrum reduces the theoretical uncertainty of the
prediction for the total branching ratio and hence the extracted value
of $|V_{ub}|$.
In order to constrain $\phi_\pi$ even further,
it will be necessary to perform a combined analysis including also
other experimental constraints from e.g.\ the $\pi$ electromagnetic form
factor \cite{piDA} and the $\pi$--$\gamma$ transition form factor
\cite{stefanis}. 

The determination of $|V_{ub}|$ 
presented in this letter can be improved in the future 
from both the experimental and the theoretical side. As for the
latter, we would like to stress that the normalisation of $f_+$ depends
on the treatment of $f_B$, the decay constant of the $B$ meson. We
have determined $f_B$ from a QCD sum rule to $O(\alpha_s)$ accuracy,
using the same values for quark mass and quark condensate as in the
LCSR for $f_Bf_+$, and we have shown, by comparison with the
corresponding tree-level sum rules, that the individually large
$O(\alpha_s)$ corrections to $f_Bf_+$ and $f_B$ cancel in the
ratio to a large extent. We have estimated the uncertainty of this
procedure to be $\sim 8\%$, which enters the normalisation of the form
factor. This overall uncertainty  can be reduced by calculating 
e.g.\ $O(\beta_0\alpha_s^2)$ corrections to the LCSRs,
which is a formidable, but not impossible task. The full
$O(\alpha_s^2)$ corrections to $f_B$ are known and actually also
reduce the uncertainty coming from $m_b$ \cite{fB}. As for the
experimental input, smaller bins in $q^2$ would help to further constrain the
shape of the form factor and ultimately avoid the necessity for
extrapolation in $q^2$.

\section*{Acknowledgements}
R.Z.\ is supported by the Swiss National Science Foundation.
P.B.\ would like to thank K.~Hamilton for enlightening discussions on
the art of error correlation \& propagation.


\begin{thebibliography}{99}

\bibitem{data} 
B.~Aubert {\it et al.}  [BABAR Collaboration],
arXiv:hep-ex/0507003.
%%CITATION = HEP-EX 0507003;%%

\bibitem{BZ04}
P.~Ball and R.~Zwicky,
%``New results on B $\to$ pi, K, eta decay formfactors from light-cone sum
%rules,''
Phys.\ Rev.\ D {\bf 71} (2005) 014015
[arXiv:hep-ph/0406232].
%%CITATION = HEP-PH 0406232;%%

\bibitem{QM}
M.~Wirbel, B.~Stech and M.~Bauer,
%``Exclusive Semileptonic Decays Of Heavy Mesons,''
Z.\ Phys.\ C {\bf 29} (1985) 637;\\
%%CITATION = ZEPYA,C29,637;%%
J.~G.~K\"orner and G.~A.~Schuler,
%``Exclusive Semileptonic Decays Of Bottom Mesons In The Spectator Quark
%Model,''
Z.\ Phys.\ C {\bf 38} (1988) 511
[Erratum-ibid.\ C {\bf 41} (1989) 690];\\
%%CITATION = ZEPYA,C38,511;%%
N.~Isgur {\it et al.},
%``Semileptonic B And D Decays In The Quark Model,''
Phys.\ Rev.\ D {\bf 39} (1989) 799.
%%CITATION = PHRVA,D39,799;%%

\bibitem{BZ01}
P.~Ball,
%``B $\to$ pi and B $\to$ K transitions from {QCD} sum rules on the
%light-cone,''
JHEP {\bf 9809} (1998) 005
[arXiv:hep-ph/9802394];\\
%%CITATION = HEP-PH 9802394;%%
P.~Ball and R.~Zwicky,
%``Improved analysis of B $\to$ pi e nu from QCD sum rules on the light-cone,''
JHEP {\bf 0110} (2001) 019
[arXiv:hep-ph/0110115].
%%CITATION = HEP-PH 0110115;%%

\bibitem{otherLCSR}
V.M.\ Belyaev, A. Khodjamirian and R. R\"uckl,
%``QCD calculation of the B $\to$ pi, K form-factors,''
Z.\ Phys.\ C {\bf 60} (1993) 349
[hep-ph/9305348];\\
%%CITATION = HEP-PH 9305348;%%
A. Khodjamirian {\it et al.},
%``Perturbative QCD correction to the B $\to$ pi transition form-factor,''
Phys.\ Lett.\ B {\bf 410} (1997) 275
[hep-ph/9706303];\\
%%CITATION = HEP-PH 9706303;%%
E. Bagan, P. Ball and V.M.\ Braun,
%``Radiative corrections to the decay B $\to$ pi e nu and the heavy quark
Phys.\ Lett.\ B {\bf 417} (1998) 154
[hep-ph/9709243];\\
%%CITATION = HEP-PH 9709243;%%
A. Khodjamirian {\it et al.},
%``Predictions on B $\to$ pi anti-l nu/l, D $\to$ pi anti-l nu/l and D $\to$
Phys.\ Rev.\ D {\bf 62} 114002 (2000)
[hep-ph/0001297].
%%CITATION = HEP-PH 0001297;%%

\bibitem{lattice}
J.~Shigemitsu {\it et al.},
%``Semileptonic B decays with N(f) = 2+1 dynamical quarks,''
arXiv:hep-lat/0408019;\\
%%CITATION = HEP-LAT 0408019;%%
M.~Okamoto {\it et al.},
%``Semileptonic D $\to$ pi / K and B $\to$ pi / D decays in 2+1 flavor lattice
%QCD,''
Nucl.\ Phys.\ Proc.\ Suppl.\  {\bf 140} (2005) 461
[arXiv:hep-lat/0409116].
%%CITATION = HEP-LAT 0409116;%%

\bibitem{BK}
D.~Becirevic and A.~B.~Kaidalov,
%``Comment on the heavy $\to$ light form factors,''
Phys.\ Lett.\ B {\bf 478}, 417 (2000)
[arXiv:hep-ph/9904490].
%%CITATION = HEP-PH 9904490;%%

\bibitem{dispersion}
L.~Lellouch,
%``Lattice-Constrained Unitarity Bounds for $\bar B~0\to\pi+\ell-\bar\nu_\ell$
%Decays,''
Nucl.\ Phys.\ B {\bf 479} (1996) 353
[arXiv:hep-ph/9509358];\\
%%CITATION = HEP-PH 9509358;%%
M.~Fukunaga and T.~Onogi,
%``A model independent determination of $|$V(ub)$|$ using the global q**2
%dependence of the dispersive bounds on the B $\to$ pi l nu form factors,''
Phys.\ Rev.\ D {\bf 71} (2005) 034506
[arXiv:hep-lat/0408037];\\
%%CITATION = HEP-LAT 0408037;%%
M.~C.~Arnesen {\it et al.},
%``A precision model independent determination of $|$V(ub)$|$ from B $\to$ pi e
%nu,''
arXiv:hep-ph/0504209.
%%CITATION = HEP-PH 0504209;%%

\bibitem{prevdata}
S.~B.~Athar {\it et al.}  [CLEO Collaboration],
%``Study of the q**2 dependence of B $\to$ pi l nu and B $\to$ rho(omega) l  nu
%decay and extraction of $|$V(ub)$|$,''
Phys.\ Rev.\ D {\bf 68} (2003) 072003
[arXiv:hep-ex/0304019];\\
%%CITATION = HEP-EX 0304019;%%
K.~Abe {\it et al.}  [BELLE Collaboration],
  %``Measurement of exclusive B $\to$ X/u l nu decays with D(*) l nu decay
  %tagging,''
  arXiv:hep-ex/0408145;\\
  %%CITATION = HEP-EX 0408145;%%
B.~Aubert  [BABAR Collaboration],
  %``Branching fraction for B0 $\to$ pi- l+ nu and determination of $|$V(ub)$|$
  %in Upsilon(4S) $\to$ B0 anti-B0 events tagged by anti-B0 $\to$ D(*)+ l-
  %anti-nu,''
  arXiv:hep-ex/0506064.
  %%CITATION = HEP-EX 0506064;%%

\bibitem{talbot}
P.~Ball and A.~N.~Talbot,
%``Models for light-cone meson distribution amplitudes,''
JHEP {\bf 0506} (2005) 063 [arXiv:hep-ph/0502115].
%%CITATION = HEP-PH 0502115;%%

\bibitem{BF}
V.~M.~Braun and I.~E.~Filyanov,
%``QCD Sum Rules In Exclusive Kinematics And Pion Wave Function,''
Z.\ Phys.\ C {\bf 44}, 157 (1989)
[Sov.\ J.\ Nucl.\ Phys.\  {\bf 50}, 511 (1989)].
%%CITATION = ZEPYA,C44,157;%%

\bibitem{piDA}
S.~V.~Mikhailov and A.~V.~Radyushkin,
%``The Pion wave function and QCD sum rules with nonlocal condensates,''
Phys.\ Rev.\ D {\bf 45} (1992) 1754;\\
%%CITATION = PHRVA,D45,1754;%%                                           
A.~Schmedding and O.~I.~Yakovlev,
%``Perturbative effects in the form factor gamma gamma* $\to$ pi0 and
%extraction of the pion wave function from CLEO data,''
Phys.\ Rev.\ D {\bf 62} (2000) 116002
[arXiv:hep-ph/9905392];\\
%%CITATION = HEP-PH 9905392;%%
V.~M.~Braun, A.~Khodjamirian and M.~Maul,
%``Pion form factor in QCD at intermediate momentum transfers,''
Phys.\ Rev.\ D {\bf 61} (2000) 073004
[arXiv:hep-ph/9907495].
%%CITATION = HEP-PH 9907495;%%

\bibitem{stefanis}
A.~P.~Bakulev, S.~V.~Mikhailov and N.~G.~Stefanis,
%``QCD-based pion distribution amplitudes confronting experimental data,''
Phys.\ Lett.\ B {\bf 508} (2001) 279
[Erratum-ibid.\ B {\bf 590} (2004) 309]
[arXiv:hep-ph/0103119];\\
%%CITATION = HEP-PH 0103119;%%
A.~P.~Bakulev {\em et al.},
%``Pion form factor in QCD: From nonlocal condensates to NLO analytic
%perturbation theory,''
Phys.\ Rev.\ D {\bf 70} (2004) 033014
[Erratum-ibid.\ D {\bf 70} (2004) 079906]
[arXiv:hep-ph/0405062].
%%CITATION = HEP-PH 0405062;%%

\bibitem{fB_latt}
S.~Hashimoto,
%``Recent results from lattice calculations,''
arXiv:hep-ph/0411126.
%%CITATION = HEP-PH 0411126;%%

\bibitem{HFAG}
Heavy Flavor Averaging Group (HFAG) (J. Alexander {\it et al.}),\\ 
{\tt http://www.slac.stanford.edu/xorg/hfag/semi/lp05/update.html}

\bibitem{fB}
A.~A.~Penin and M.~Steinhauser,
%``Heavy-light meson decay constant from QCD sum rules in three-loop
%approximation,''
Phys.\ Rev.\ D {\bf 65}, 054006 (2002)
[arXiv:hep-ph/0108110];\\
%%CITATION = HEP-PH 0108110;%%
M.~Jamin and B.~O.~Lange,
%``f(B) and f(B/s) from QCD sum rules,''
Phys.\ Rev.\ D {\bf 65}, 056005 (2002)
[arXiv:hep-ph/0108135].
%%CITATION = HEP-PH 0108135;%%

\end{thebibliography}
\end{document}